\documentclass[fleqn,usenatbib]{mnras}

\usepackage{newtxtext,newtxmath}

\usepackage[T1]{fontenc}
\usepackage{ae,aecompl}

\usepackage{graphicx}	
\usepackage{amsmath}	
\usepackage{amssymb}	

\newcommand{\nustar}{\textit{NuSTAR }}

\newcommand{\swift}{{\it Swift }}
\newcommand{\xmm}{{\it XMM-Newton }}

\newcommand{\blue}{\textcolor{blue}}

\title[High Density Reflection I.]{High Density Reflection Spectroscopy I. A case study of GX~339-4}

\author[J. Jiang et al.]{
Jiachen Jiang,$^{1}$\thanks{E-mail: jj447@cam.ac.uk} Andrew C. Fabian,$^{1}$ Jingyi Wang,$^{2}$ Dominic J. Walton,$^{1}$  
\newauthor Javier A. Garc{\'{\i}}a,$^{3,4}$ Michael L. Parker$^{5}$, James F. Steiner,$^{2}$ and John A. Tomsick$^{6}$
\\
$^{1}$Institute of Astronomy, University of Cambridge, Madingley Road, Cambridge CB3 0HA\\
$^{2}$MIT Kavli Institute for Astrophysics and Space Research, MIT, 70 Vassar Street, Cambridge, MA 02139, USA\\
$^{3}$Cahill Center for Astronomy and Astrophysics, California Institute of Technology, Pasadena, CA 91125, USA\\ 
$^{4}$Dr. Karl Remeis-Observatory and Erlangen Centre for Astroparticle Physics, Sternwartstr. 7, D-96049 Bamberg, Germany\\
$^{5}$European Space Agency (ESA), European Space Astronomy Centre (ESAC), E-28691 Villanueva de la Ca\~nada, Spain\\
$^{6}$Space Sciences Laboratory, 7 Gauss Way, University of California, Berkeley, CA 94720-7450, USA\\
}

\date{Accepted XXX. Received YYY; in original form ZZZ}

\pubyear{2018}

\begin{document}
\label{firstpage}
\pagerange{\pageref{firstpage}--\pageref{lastpage}}
\maketitle

\begin{abstract}
We present a broad band spectral analysis of the black hole binary GX~339-4 with \nustar and \swift using high density reflection model. The observations were taken when the source was in low flux hard states (LF) during the outbursts in 2013 and 2015, and in a very high flux soft state (HF) in 2015. The high density reflection model can explain its LF spectra with no requirement for an additional low temperature thermal component. This model enables us to constrain the density in the disc surface of GX~339-4 in different flux states. The disc density in the LF state is $\log(n_{\rm e}/$\,cm$^{-3})\approx21$, 100 times higher than the density in the HF state ($\log(n_{\rm e}/$\,cm$^{-3})=18.93^{+0.12}_{-0.16}$). A close-to-solar iron abundance is obtained by modelling the LF and HF broad band spectra with variable density reflection model ($Z_{\rm Fe}=1.50^{+0.12}_{-0.04}Z_{\odot}$ and $Z_{\rm Fe}=1.05^{+0.17}_{-0.15}Z_{\odot}$ respectively).
\end{abstract}




\section{Introduction}

The primary X-ray spectra from black holes (BHs) can be described by a power-law continuum, which originates from a high temperature compact structure external to the black hole accretion disc. This high temperature compact structure is called the corona. The interaction between the primary power-law photons and the disc top layer can produce both emission, including fluorescence lines and recombination continuum, and absorption edges. These features are referred to as the disc reflection spectrum \citep[e.g.][]{george91,garcia10}. The disc reflection spectrum is highly affected by relativistic effects, such as Doppler effect and gravitational redshift, due to the strong gravitational field in the vicinity of black holes \citep[e.g.][]{reynolds03}. For example, relativistic blurred Fe K$\alpha$ emission line features have been detected in reflection spectra of both Active Galactic Nuclei \citep[AGN, e.g. MCG-6$-$30$-$15,][]{tanaka95} and Galactic BH X-ray Binary sources \citep[XRB, e.g. Cyg X-1,][]{barr85}. Relativistic reflection spectra can provide information on the disc-corona geometry, such as the coronal region size and the disc inner radius. By comparing relativistic reflection spectra in different observations, we can investigate changes of the inner accretion processes through the evolution of the X-ray flux states in both highly variable AGNs \citep[e.g. Mrk~335, IRAS~13224$-$3809,][]{parker14,jiang18} and XRBs that show different flux states \citep[e.g. XTE~J1650-500,][]{reis15}. 

The existence of two different flux states in XRB was first realized in the X-ray emission of the XRB Cyg X-1 \citep{oda71}. Its X-ray spectrum can change from a soft spectrum featured by a strong thermal component to a hard spectrum featured by a strong disc reflection component. The soft state, which is also characterised by no radio detection, is identified as the `high' flux (HF) state and the hard state with associated radio detection, is identified as the `low' flux (LF) state, due to the large flux variation during the transition. Measurements in the HF soft states of BH XRB offer good evidence that the accretion disc is extended to the innermost stable circular orbit \citep[ISCO, e.g. LMC X-3,][]{steiner10}. Most of the spin measurements of soft states are based on the assumption that the inner radius is located at ISCO \citep[e.g.][]{gou14,walton16}. In the LF hard state, the disc is predicted to be truncated at a large radius and replaced by an advective flow at small radii \citep{esin97,narayan05}. Although there is evidence that the disc is truncated as measured by reflection spectroscopy at X-ray luminosities $L_{\rm X}\approx0.1\% L_{\rm Edd}$ \citep{tomsick09,narayan08}, there is a substantial debate whether the disc is truncated in the intermediate flux hard state due to different spectral modelling or instrumental pile-up effects \citep[see the discussion in][]{jingyi18}.  

A common result obtained by reflection modelling of black hole X-ray spectra is high iron abundance compared to solar. For example, \citet{walton16} found a value of $Z_{\rm Fe} \approx 4 Z_{\odot}$ in Cyg X-1 and \citet{parker15} obtained $Z_{\rm Fe} \approx 4.7 Z_{\odot}$ in the same source. Similarly, an iron abundance of $Z_{\rm Fe} \approx 2-5Z_{\odot}$ is required for another BH XRB V404 Cyg \citep{walton17}. Such a high iron abundance has been commonly seen in AGNs as well \citep[e.g.][]{chiang15, parker18}. \citet{wang12} found that the metallicty of the outflows in different quasars can vary between 1.7--6.9$Z_{\odot}$. \citet{reynolds12} suggested that the radiation-pressure dominance of the inner disc may enhance the iron abundances. However radiative levitation effects make predictions for a change of the inner disc iron abundance, which is difficult to be observed in AGNs due to their longer dynamical timescales. 

Another possible explanation for the high iron abundances is high density reflection. Most versions of available disc reflection models assume a constant electron density $n_{\rm e}=10^{15}$cm$^{-3}$ for the top layer of the BH accretion disc, which is appropriate for very massive supermassive black holes in AGNs (e.g. $M_{\rm BH}>10^{8}M_{\odot}$). For example, an upper limit of $n_{\rm e}<10^{15.3}$cm$^{-3}$ is obtained in Seyfert 1 galaxy 1H0419$-$577 \citep[$M_{\rm BH}\approx1.3\times10^{8}M_{\odot}$,][]{grupe10} by fitting its \xmm spectra with variable density reflection model (Jiang submitted). At higher electron density, the free-free process becomes more important in constraining low energy photons, increasing the temperature of the top layer of the disc, and thus turning the reflected emissions below 1~keV into a blackbody shaped spectrum \citep{ross07,garcia16}. Such a model can potentially relieve the very high iron abundance required in previous reflection spectral modelling. For instance, \citet{tomsick18} obtained an electron density of $n_{\rm e}\approx3\times10^{20}$cm$^{-3}$ by fitting the Cyg X-1 intermediate flux state spectra with the high electron density reflection model. Although the iron abundance was fixed at the solar value during the spectral fitting, the model successfully explains the spectra. \citet{jiang18} fitted the narrow line Seyfert 1 galaxy IRAS~13224$-$3809 spectra and obtained an electron density of $n_{\rm e}>10^{19.7}$cm$^{-3}$ with an iron abundance of $Z_{\rm Fe}\approx5Z_{\odot}$, which is significantly lower than the previous results $Z_{\rm Fe}\approx20Z_{\odot}$ and closer to the iron abundance measured in the ultra-fast outflow of the same source.  

Higher densities may also potentially explain the weak low temperature thermal component found in the LF state of the XRBs \citep[e.g.][]{reis08,jingyi18} and at least some of the soft excess commonly seen in Seyfert galaxies \citep[e.g.][]{fabian09,chiang15,jiang18}. The inclusion of the high electron density effects significantly decreases the flux of the best-fit blackbody component in IRAS~13224$-$3809 required for the spectral fitting purpose \citep{jiang18}. It is also interesting to note that the best-fit flux and temperature of the blackbody component that accounts for the soft excess in IRAS~13224$-$3809 show a $F\propto T^{4}$ relation, indicating a constant area origin of the soft excess emission \citep{chiang15,jiang18}.

GX~339-4 is a low mass X-ray binary (LMXB) and shows activity in a wide range of wavelength from optical to X-ray. The mass of the central black hole still remains uncertain. For example, \citet{heida17} obtained a black hole mass of $2-10M_{\odot}$ by studying its near infrared spectrum and a mass of $>5M_{\odot}$ is obtained previously by \citet{hynes03, hynes05,munoz08}. The distance has been estimated to be $\approx7$\,kpc \citep{zdziarski04}. GX~339$-4$ has shown frequent outbursts and multiple X-ray observations have been taken during different spectral states of GX~339-4. In its hard state, its X-ray spectrum shows a broad iron emission line and a power-law continuum with a photon index varying between $\Gamma\approx1.5-2.5$ across different flux levels \citep{miller04,miller06,miller08}. \citet{reis08} presented a systematic study of its high and low hard state \xmm and \textit{RXTE} spectra by taking the blackbody radiation from the disc into the top layer as well as the Comptonization effects into modelling, and obtained a black hole spin of $a_*=0.94\pm0.02$. More recently, \citet{parker16} obtained a disc iron abundance of $Z_{\rm Fe} \approx 6.6Z_{\odot}$ for the HF soft state \nustar and \swift spectra of GX~339-4. In this study, the disc inner radius is assumed to be located at ISCO and a black hole spin of $a_*>0.95$ is obtained by combining disc thermal spectral and reflection spectral modelling. Later, \citet{jingyi18} found $Z_{\rm Fe} \approx 8 Z_{\odot}$ for the LF state of the same source observed by the same instruments. Similar conclusions were found by analysing its stacked \textit{RXTE} spectra at the LF states \citep{garcia15} and \nustar spectra during the outburst of 2013 \citep{furst15}.

In this paper, we present a high density reflection interpretation of both LF and HF state spectra of GX~339-4. The same \nustar and \swift spectra as in \citet{parker16,jingyi18} are considered. In Section \ref{reduction}, we introduce the data reduction process; in Section \ref{spectrum}, we introduce the details of high density reflection modelling of the LF and HF spectra of GX~339-4; in Section \ref{results_spectrum}, we present and discuss the final spectral fitting results. The high density reflection modelling of AGN spectra are presented in a companion paper (Jiang in prep).

\section{Observations and Data Reduction} \label{reduction}

\begin{figure}
	\includegraphics[width=\columnwidth]{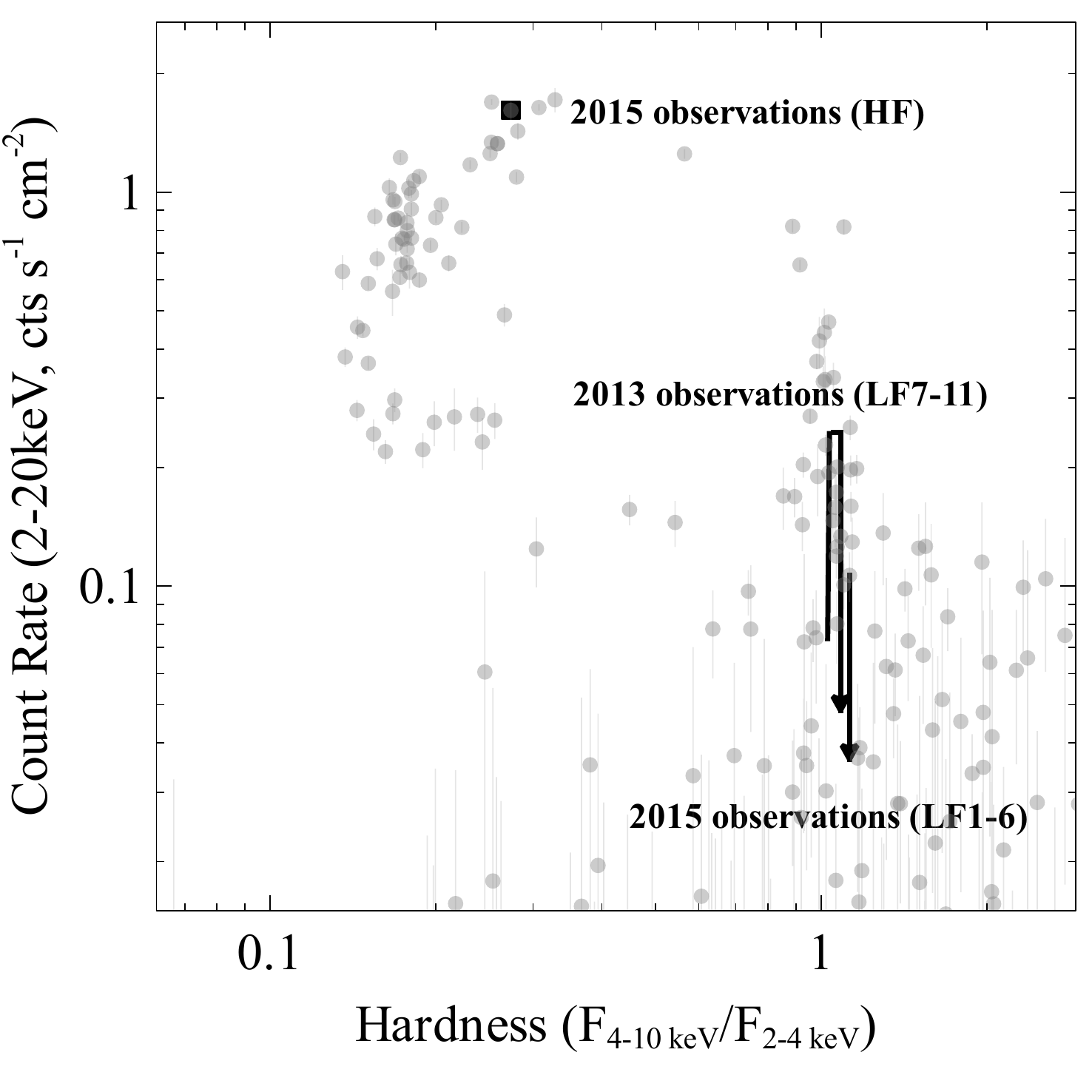}
    \caption{Weekly \textit{MAXI} hardness-intensity diagram for the 2009-2018 period of GX~339-4. The black square and the arrows correspond to the dates of the HF observations and the LF(1-11) observations analysed in this work. The arrows show the flux change during the \nustar monitoring of the outbursts. \nustar observations were taken during the rise and decay of the outburst in 2013, and only during the decay of the outburst in 2015.}
    \label{pic_hi}
\end{figure}

\begin{table*}
\caption{\nustar and \swift observations of GX~339-4 in 2013 and 2015. WT: window timing mode; PC: photon counting mode.}
\begin{tabular}{cccccccc}
\hline\hline
Obs & \nustar obsID & Date & exp.(ks) & \swift obsID & Date & exp.(ks) & Mode \\ 
\hline
HF  & 80001015003 & 2015-03-04 & 30.9 & 00081429002 & 2015-03-04 & 1.9 & WT \\
LF1 & 80102011002 & 2015-08-28 & 21.6 & 00032898124 & 2015-08-29 & 1.7 & WT \\
LF2 & 80102011004 & 2015-09-02 & 18.3 & 00032898126 & 2015-09-03 & 2.3 & WT \\
LF3 & 80102011006 & 2015-09-07 & 19.8 & 00032898130 & 2015-09-07 & 2.8 & WT \\
LF4 & 80102011008 & 2015-09-12 & 21.5 & 00081534001 & 2015-09-12 & 2.0 & PC \\
LF5 & 80102011010 & 2015-09-17 & 38.5 & 00032898138 & 2015-09-17 & 2.3 & WT \\
LF6 & 80102011012 & 2015-09-30 & 41.3 & 00081534005 & 2015-09-30 & 2.0 & PC \\
\hline
LF7 & 80001013002 & 2013-08-11 & 42.3 & 00032490015 & 2013-08-12 & 1.1 & WT \\
LF8 & 80001013004 & 2013-08-16 & 47.4 & 00080180001 & 2013-08-16 & 1.9 & WT \\
LF9 & 80001013006 & 2013-08-24 & 43.4 & 00080180002 & 2013-08-24 & 1.6 & WT \\
LF10& 80001013008 & 2013-09-03 & 61.9 & 00032898013 & 2013-09-02 & 2.0 & WT \\
LF11& 80001013010 & 2013-10-16 & 98.2 & 00032988001 & 2013-10-17 & 9.6 & WT \\
\hline\hline
\end{tabular}
\label{tab_obs}
\end{table*}

The weekly \textit{MAXI} hardness-intensity diagram (HID) for the 2009-2018 period of GX~339-4 \citep{matsuoka09} is shown in Fig.\,\ref{pic_hi}, showing a standard `q-shaped' behaviour during the outbursts. GX~339-4 went through two outbursts each in 2013 and 2015. 11 \nustar observations in total, each with a corresponding \swift snapshot, were triggered during these two outbursts, shown by the arrow in Fig.\,\ref{pic_hi}. The \nustar LF observations in 2015 were taken only during the decay of the outburst. In this work, we consider all of the \nustar observations taken during these two outbursts. In March 2015, GX~339-4 was detected with strong thermal and power-law components by \textit{Swift}, suggesting strong evidence of a HF state with a combination of disc thermal component and reflection component. One \nustar target of opportunity observation was triggered with a simultaneous \swift snapshot. See the black square in Fig.\,\ref{pic_hi} for the flux and hardness state of the source during its HF observations.  A full list of observations are shown in Table \ref{tab_obs}. 
 
\subsection{\nustar Data Reduction}

The standard pipeline NUPIPELINE V0.4.6, part of HEASOFT V6.23 package, is used to reduce the \nustar data. The \nustar calibration version V20171002 is used. We extract source spectra from circular regions with radii of 100\,arcsec, and the background spectra from nearby circular regions on the same chip. The task NUPRODUCTS is used for this purpose. The 3-78\,keV band is considered for both FPMA and FPMB spectra. The spectra are grouped to have a minimum signal-to-noise (S/N) of 6 and to oversample by a factor of 3.

\subsection{\swift Data Reduction}

The \swift observations are processed using the standard pipeline XRTPIPELINE V0.13.3. The calibration file version used is x20171113. The LF observations taken in the WT mode are not affected by the pile-up effects. The source spectra are extracted from a circular region with a radius of 20 pixels \footnote{1 pixel $\approx2.36''$} and the background spectrum spectra are extracted from an annular region with an inner radius of 90 pixels and an outer radius of 100 pixels. The LF observations taken in the PC mode are affected by the pile-up effects. By following \citet{jingyi18} where they estimated the PSF file, a circular region with a radius of 5 pixels is excluded in the center of the source region. The 0.6--6~keV band of all the LF \swift XRT spectra are considered. The HF observation was taken in the WT mode and was affected by pile-up effects. By following \citet{parker16}, a circular radius of 10 pixels is excluded in the center of the source region. The 0.6--1~keV of the HF \swift XRT spectrum at a very high flux state is ignored due to known issues of the RMF redistribution issues in the WT mode \footnote{See following website for more XRT WT mode calibration information. http://www.swift.ac.uk/analysis/xrt/digest\_cal.php}. The \swift XRT spectra are grouped to have a minimum S/N of 6 and to oversample by a factor of 3.

\section{Spectral Analysis} \label{spectrum}

XSPEC V12.10.0.C \citep{arnaud96} is used for spectral analysis, and C-stat is considered in this work. The Galactic column density towards GX~339-4 remains uncertain. The value of combined $N_{\rm H_I}$ and $N_{\rm H_2}$ obtained by \citet{willingale13} is $5.18\times10^{21}$cm$^{-2}$. However, \citet{kalberla05} reported a column density of $3.74\times10^{21}$cm$^{-2}$ in the Leiden/Argentine/Bonn survey. The Galactic column density values measured by different sets of broad band X-ray spectra are different too. For example, \citet{jingyi18} obtained $\approx4\times10^{21}$cm$^{-2}$ while \citet{parker16} obtained a higher value of $7.7\pm0.2\times10^{21}$cm$^{-2}$. We therefore fixed the Galactic column density at $3.74\times10^{21}$cm$^{-2}$ in the beginning of our analysis and allow it to vary to obtain the best-fit value for each set of spectra. For local Galactic absorption, the \texttt{tbabs} model is used. The solar abundances of \citet{wilms00} are used in \texttt{tbabs}. An additional constant model \texttt{constant} has been applied to vary normalizations between the simultaneous spectra obtained by different instruments to account for calibration uncertainties.

\begin{figure*}
	\includegraphics[width=18cm]{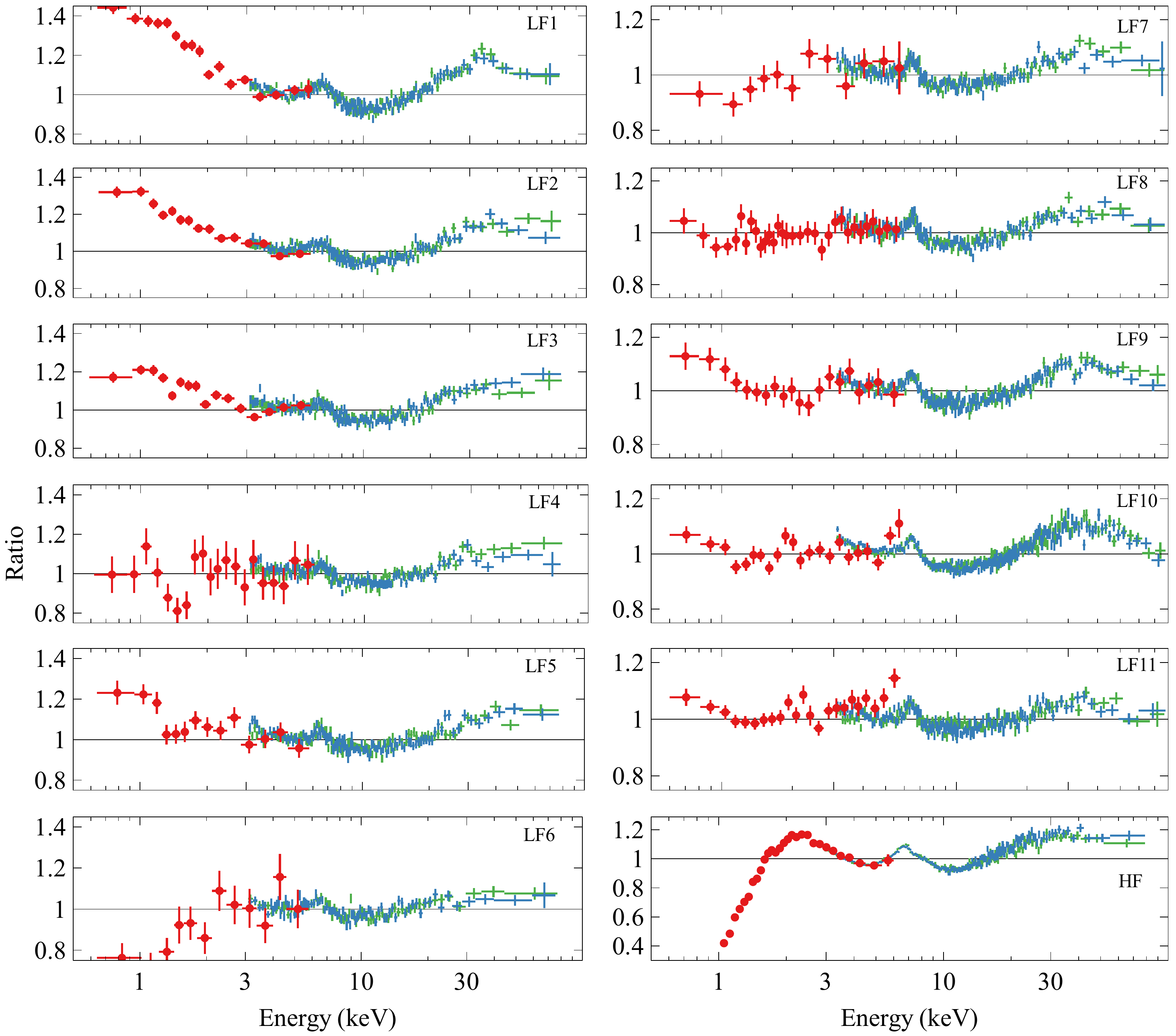}
    \caption{First 11 panels: ratio plots of GX~339-4 FPM (blue crosses: FPMA; green crosses: FPMB) and XRT (red circles) spectra fitted with a Galactic absorbed power law for LF observations in 2015 (LF1-6) and 2013 (LF7-11). Last panel: ratio plot of HF spectra fitted with a Galactic absorbed power law plus a simple blackbody component for the very high flux soft state observation in 2015. All the spectra show a broad line emission feature around 6.4~keV and a strong Compton hump above 10~keV.}
    \label{pic_fe}
\end{figure*}


\subsection{Low Flux State (LF) Spectral Modelling} \label{lf_spectrum}

\begin{figure}
	\includegraphics[width=\columnwidth]{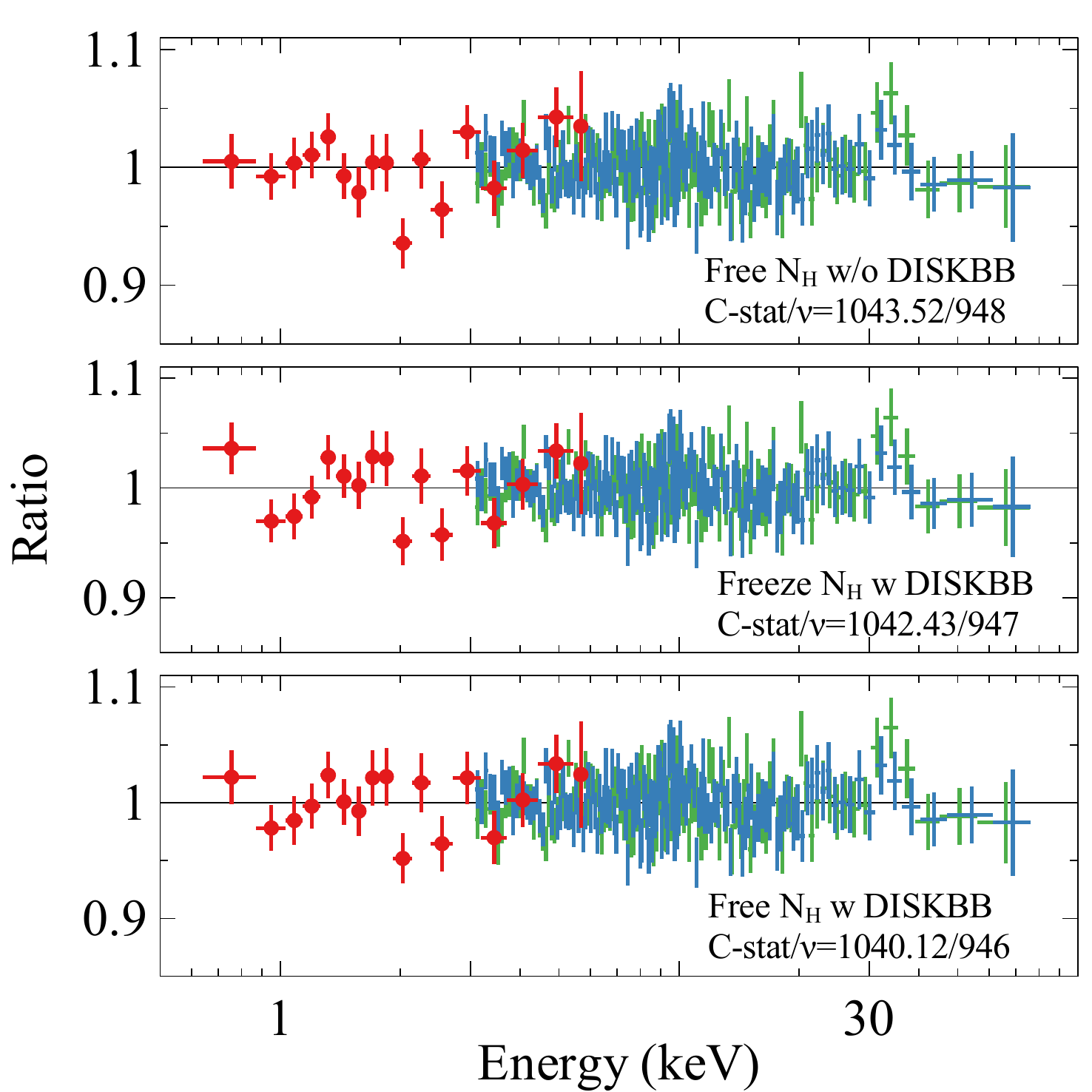}
    \caption{Ratio plots for LF1 spectra against different continuum models. Red circles: XRT; blue crosses: FPMA; green crosses: FPMB. See text for more details.}
    \label{pic_lf1}
\end{figure}

We analyze all the LF \nustar observations publicly available prior to 2018 and they have discussed in \citet{furst15,jingyi18}. Fig.\,\ref{pic_fe} shows the ratio plots of LF1-11 spectra fitted with a Galactic absorbed power-law model obtained by fitting only the corresponding \nustar spectra. All the LF spectra show a broad emission line feature around 6.4~keV with a Compton hump above 20~keV. They provide a strong evidence of a relativistic disc reflection component. By following \citet{garcia15,jingyi18}, we model the features with a combination of relativistic disc reflection and a distant reflector for the narrow emission line component. A more developed version of \texttt{reflionx} \citep{ross05} is used to model the rest-frame disc reflection spectrum. The \texttt{reflionx} grid allows the following free parameters: disc iron abundance ($Z_{\rm Fe}$), disc ionization $\log(\xi)$, disc electron density $n_{\rm e}$, high energy cutoff ($E_{\rm cut}$), and photon index ($\Gamma$). All the other element abundances are fixed at the solar value \citep{morrison83}. The ionization parameter is defined as $\xi=4\pi F/n$, where $F$ is the total illuminating flux and $n$ is the hydrogen number density. The photon index $\Gamma$ and high energy cutoff $E_{\rm cut}$ are linked to the corresponding parameters of the coronal emission modelled by \texttt{cutoffpl} in XSPEC. A convolution model \texttt{relconv} \citep{dauser13} is applied to the rest frame ionized disc reflection model \texttt{reflionx} to apply relativistic effects. A simple power-law shaped emissivity profile is assumed ($\epsilon\propto r^{-q}$) and the emissvity index $q$ is allowed to vary during the fit. Other free parameters are the disc viewing angle $i$ and the disc inner radius $r_{\rm in}$/ISCO. The ionization of the distant reflector is fixed at the minimum value $\xi=10$. The other parameters of the distant reflector are linked to the corresponding parameters in the disc reflection component. The BH spin parameter $a_*$ is fixed at its maximum value $0.998$ \citep{kerr63} to fully explore the $r_{\rm in}$ parameter. We use \texttt{cflux}, a simple convolution model in XSPEC, to calculate the 1--10~keV flux of each model component. For future reference and simplicity, we define an empirical reflection fraction as $f_{\rm refl}=F_{\rm ref}/F_{\rm pl}$ in the 1--10~keV band, where $F_{\rm ref}$ and $F_{\rm pl}$ are the flux of the disc reflection component and the coronal emission calculated by \texttt{cflux}. Note that this is not the same as the physically defined reflection fraction discussed by \citet{dauser16}. The final model is \texttt{tbabs * ( cflux*(relconv*reflionx) + cflux*reflionx + cflux*cutoffpl)} in XSPEC format. This model can fit all LF spectra successfully with no obvious residuals. For example, it offers a good fit for the LF1 spectra with C-stat/$\nu$ = 1043.52/948. A ratio plot of LF1 spectra fitted with this model is shown in the top panel of Fig.\,\ref{pic_lf1}. The best-fit values of some key parameters that affect the spectral modelling below 3~keV are following: $N_{\rm H}=3.4^{+0.2}_{-0.1}\times10^{21}$\,cm$^{-2}$, $\log(\xi$/erg cm s$^{-1})=3.18^{+0.07}_{-0.06}$, and $\log(n_{\rm e}$/cm$^{-3})=20.6\pm0.3$. Our best-fit column density is consistent with the Galactic column density measured in \citet{kalberla05}.

\begin{figure}
	\includegraphics[width=\columnwidth]{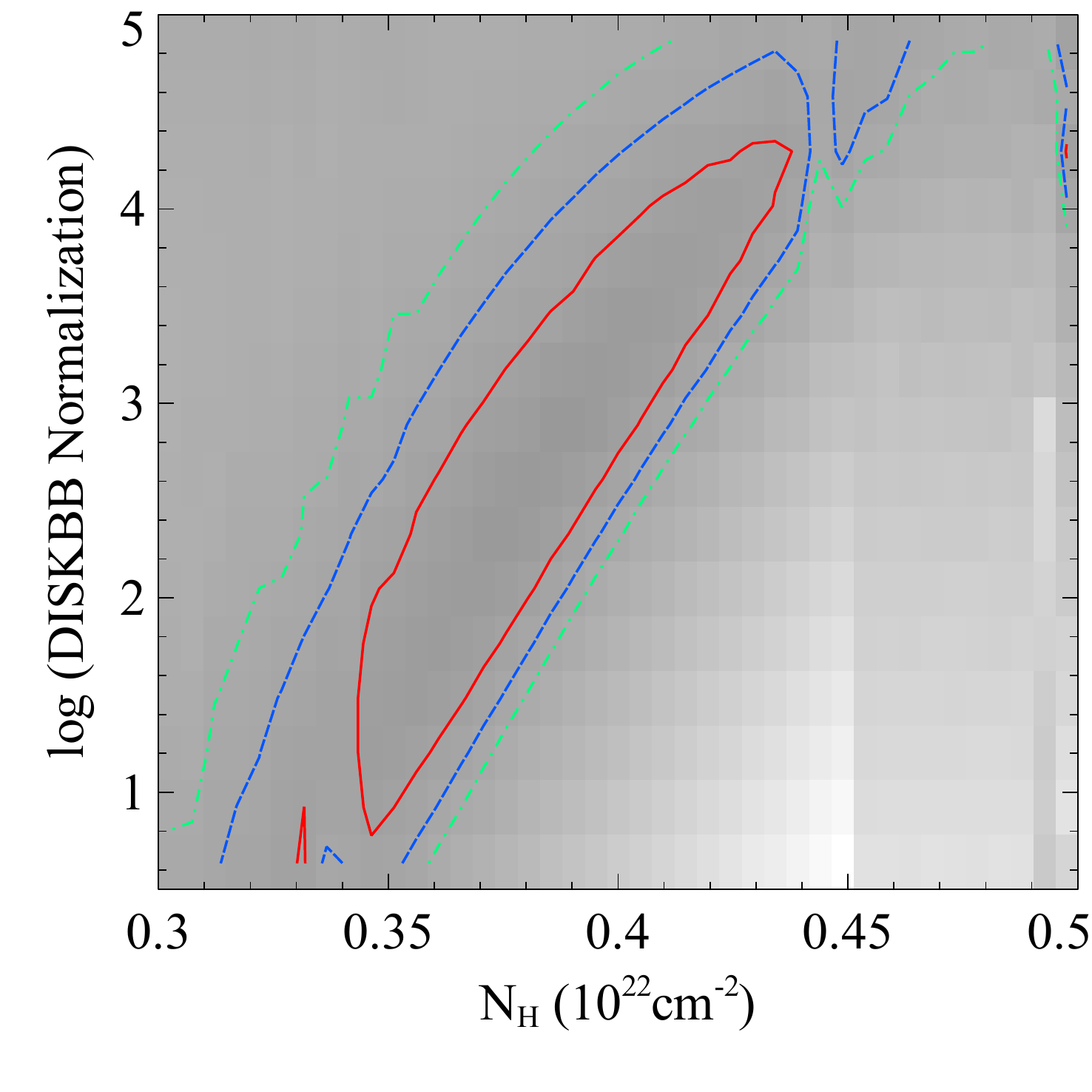}
    \caption{A contour plot of C-stat distribution on the Galactic absorption column density $N_{\rm H}$ vs. \texttt{diskbb} model normalization parameter plane for LF1 spectra when fitted with \texttt{tbabs*(diskbb+cutoffpl+relconv*reflionx+reflionx)}. It shows a clear degeneracy between two parameters. The lines show the 1$\sigma$ (red solid line), 2$\sigma$ (blue dashed line), and 3$\sigma$ contours (green dash-dot line).}
    \label{pic_contour_lf1}
\end{figure}

\begin{figure}
	\includegraphics[width=8cm]{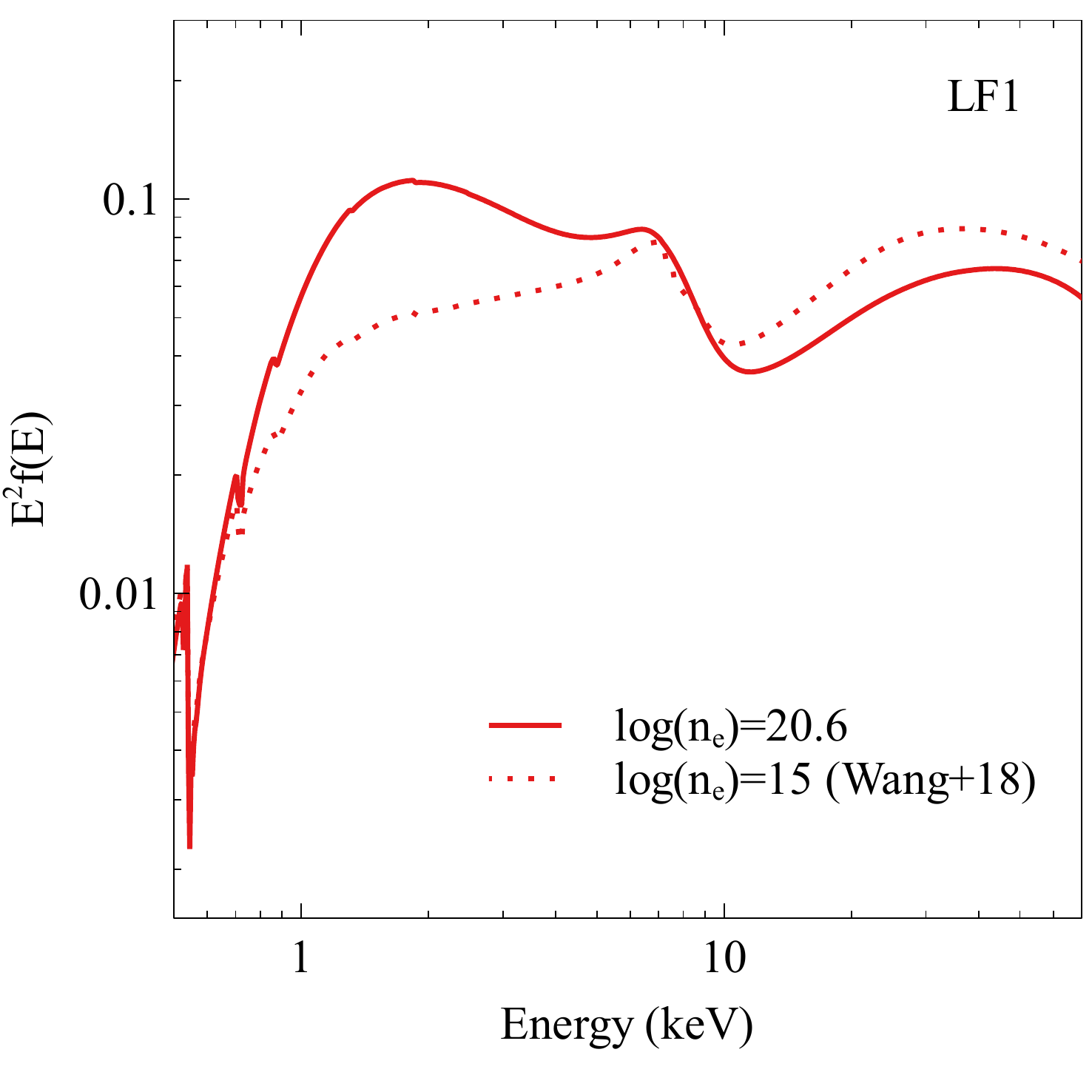}
    \caption{The best-fit relativistic high density reflection model for LF1 spectra (solid line) and the best-fit relativistic reflection model obtained by \citet{jingyi18} assuming $\log(n_{\rm e}/{\rm cm^{-3}})=15$ (dotted line). The plot is only shown for comparison of spectral shape. An additional \texttt{diskbb} component is required to fit the broad band spectra in \citet{jingyi18}.}
    \label{pic_comparison}
\end{figure}

We notice that previously the spectral modelling requires a low temperature multicolour disc thermal component \texttt{diskbb} ($kT=0.46$\,keV) when using the model with the disc electron density $n_{\rm e}$ fixed at $\log(n_{\rm e}$/cm$^{-3})=15$ for LF1 observation \citep{jingyi18}. However the normalization of this component is very low and weakly constrained. Similarly, a weak thermal component is also required in the analysis of its \xmm hard state observations \citep{reis08} and other earlier \nustar observations \citep{reis15}. The difference in spectral modelling may result from the following two reasons: one is the high density reflection model, where a blackbody-shaped emission arises in the soft band when the disc electron density $n_{\rm e}$ becomes higher than $10^{15}$\,cm$^{-3}$; the other is the uncertain neutral absorber column density, which was measured to be $N_{\rm H}=4.12^{+0.08}_{-0.12}\times10^{21}$\,cm$^{-2}$ in \citet{jingyi18} and higher than our best-fit value for the LF1 spectra. 

In order to test for an additional \texttt{diskbb} component, we first fit the spectra with $N_{\rm H}$ fixed at the higher Galactic column density $N_{\rm H}=5.18\times10^{21}$\,cm$^{-2}$ obtained by \citet{willingale13} rather than the value from \citet{kalberla05}. An additional \texttt{diskbb} component improves the fit by only $\Delta$C-stat=1.1. See the middle panel of Fig.\,\ref{pic_lf1} for the corresponding ratio plot. Only an upper limit of the \texttt{diskbb} normalization is of $N_{\rm diskbb}<1.5\times10^{5}$ found. Compared with the result in \citet{jingyi18}, a lower disc inner temperature of $kT=0.24^{+0.08}_{-0.10}$~keV is required in this fit. Second, we fit LF1 spectra with the absorber column density as a free parameter (bottom panel of Fig\,\ref{pic_lf1}). A contour plot of C-stat distribution on the $N_{\rm H}$ vs. $N_{\rm diskbb}$ parameter plane is calculated by STEPPAR function in XSPEC and shown in Fig.\,\ref{pic_contour_lf1}. It clearly shows a strong degeneracy between the absorber column density and the normalization of the \texttt{diskbb} component. The fit is only improved by $\Delta$C-stat=3 with 2 more free parameters after including this \texttt{diskbb} component. See Fig.\,\ref{pic_lf1} for ratio plots against different continuum models. By varying the Galactic column density, it only slightly changes the fit of the \swift XRT spectrum. Therefore, we conclude that an additional \texttt{diskbb} component is not necessary for LF1 spectral modelling when the disc density parameter $n_{\rm e}$ is allowed to vary. In order to visualize the spectral difference with different $n_{\rm e}$, we show the best-fit reflection model component for LF1 in Fig.\,\ref{pic_comparison} in comparison with the best-fit model for the same observation assuming $\log(n_{\rm e}/{\rm cm^{-3}})=15$\,cm$^{-3}$ in \citet{jingyi18}. With a disc density as high as $\log(n_{\rm e}/{\rm cm^{-3}})=20.6$, a quasi-blackbody emission arises in the soft band and accounts for the excess emission below 2~keV. Similar conclusions are found for the other sets of LF spectra. Future pile-up free high S/N observation below 2~keV, such as from \textit{NICER}, may help constrain more detailed spectral shape of LF states of GX~339-4.

\begin{figure*}
	\includegraphics[width=17cm]{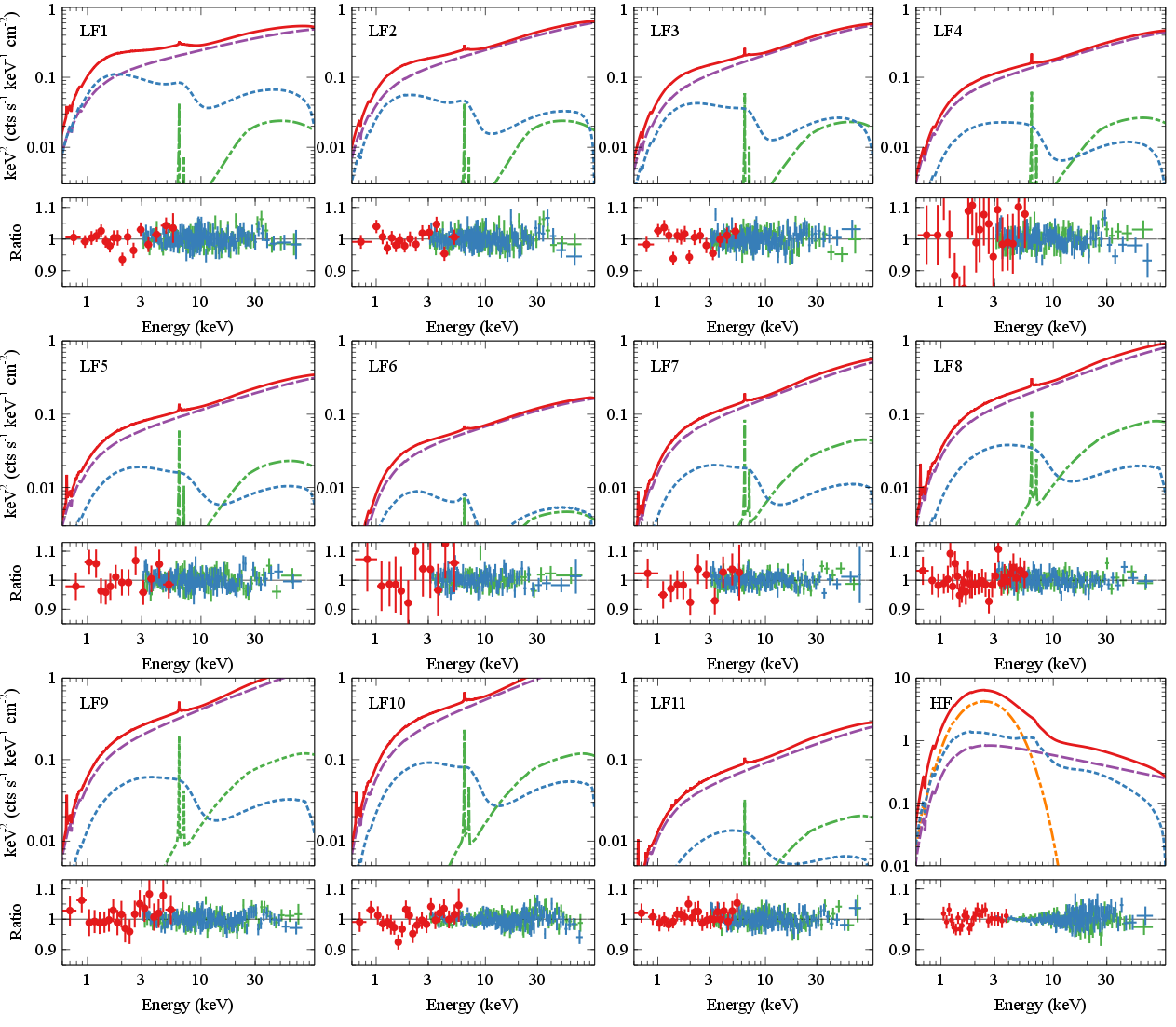}
    \caption{Top: the first 11 panels show the best-fit models obtained by fitting LF1-11 spectra simultaneously. The last panel shows the best-fit model obtained by fitting only HF spectra. Red solid lines: total model; blue dotted lines: relativistic reflection model; purple dashed lines: coronal emission; green dash-dot lines: distant reflection component; orange dash-dot-dot lines: disc thermal spectrum (only needed in the HF spectral modelling). Bottom: ratio plots against the corresponding best-fit models shown in the upper panels. Red circles: XRT; blue crosses: FPMA; green crosses: FPMB.}
    \label{pic_final}
\end{figure*}

So far we have achieved the best-fit model for the LF spectra individually. We also undertake a multi-epoch spectral analysis with disc iron abundance $Z_{\rm Fe}$ and disc viewing angle $i$ linked between LF1-11 spectra. All the other parameters are allowed to vary during the fit. A table of all the best-fit parameters are shown in Table.\,\ref{tab_fit}. The best-fit models and corresponding ratio plots are shown in Fig.\,\ref{pic_final}. We allow the column density of the neutral absorber to vary in different epochs to investigate any variance. A slightly higher column density ($N_{\rm H}\approx4.1\times10^{21}$\,cm$^{-2}$) is required for LF6,7. The emissivity index of the relativistic reflection spectrum is weakly constrained in LF3-6 observations but largely consistent with the Newtonian value $q=3$, except for the LF1 observation. We can also confirm that the disc is not truncated at a significantly large radius, such as $r=100r_g$ \citep{plant15}. A slight iron over abundance compared to solar is required ($Z_{\rm Fe}=1.5^{+0.12}_{-0.04}$) for the spectral fitting. The power-law continuum is softer in the first two observations taken at higher flux levels but remains consistent at 90\% confidence during the rest of the decay in 2015. The photon index in LF7-10 during the outburst in 2013 is consistent at 90\% confidence as well. The reflection fraction decreases with the decreasing total flux during the flux decay in 2015. This is likely caused by a receding inner disc radius at the very low flux states or a change of the coronal geometry (e.g. its height above the disc). Moreover, the multi-epoch spectral analysis of all LF observations shows tentative evidence for an anti-correlation between disc density and X-ray band flux. For example, the disc density increases from $\log(n_{\rm e}/{\rm cm^{-3}})=20.60^{+0.23}_{-0.12}$ in the highest flux state (LF1) to $\log(n_{\rm e}/{\rm cm^{-3}})=21.45^{+0.06}_{-0.13}$ in the lowest flux state (LF6). The flux level of the cold reflection component remains consistent, indicating that this emission arises from stable material at a large radius from the central black hole. We will discuss other fitting results, such as the electron density measurements, in Section \ref{results_spectrum}.

\begin{table*}
\caption{The best-fit parameters obtained by fitting 1) LF1-11 spectra simultaneously 2) only HF spectra. u: unconstrained; l: linked; f: fixed. $F_{\rm refl}$ is the flux of the relativistic disc reflection component measured between 1--10~keV; $F_{\rm pl}$ is the flux of the coronal emission measured at the same energy band; $F_{\rm dis}$ is for the distant cold reflector. The reflection fraction $f_{\rm refl}$ is defined as $F_{\rm refl}/F_{\rm pl}$ for simplicity. $L_{\rm 0.1-100keV}$ is the 0.1--100~keV band Galactic-absorption corrected luminosity calculated using the best-fit model. A black hole mass $M_{\rm BH}=10M_{\odot}$ and a distance $d=10$\,kpc are assumed.}
\begin{tabular}{cccccccc}
\hline\hline
Parameter & Unit & LF1 & LF2 & LF3 & LF4 & LF5 & LF6 \\ 
\hline
$N_{\rm H}$ & 10$^{21}$cm$^{-2}$ & $3.22^{+0.14}_{-0.08}$ & $3.20^{+0.10}_{-0.12}$ & $3.1^{+0.3}_{-0.2}$ & $3.4^{+0.5}_{-0.5}$ & $3.2^{+0.2}_{-0.3}$ & $4.1^{+0.8}_{-0.7}$  \\
\hline
q & - & $6^{+3}_{-2}$ & $2.5^{+3.6}_{-0.4}$ & $<7$ & 5(u) & 6(u) & 4(u)  \\
rin & ISCO & <4.7 & <8 & <11 & $11^{+4}_{-7}$ & $<32$ & $21^{+14}_{-12}$  \\
$\log(\xi)$ & erg cm s$^{-1}$ & $3.18^{+0.06}_{-0.39}$ & $3.13^{+0.12}_{-0.08}$ & $3.12^{+0.09}_{-0.17}$ & $3.10^{+0.17}_{-0.04}$ & $3.12^{+0.11}_{-0.08}$ & $3.08^{+0.05}_{-0.02}$  \\
$\log(n_{\rm e})$ & cm$^{-3}$ & $20.60^{+0.23}_{-0.12}$ & $20.64^{+0.16}_{-0.13}$ & $21.1^{+0.4}_{-0.2}$ & $21.49^{+0.14}_{-0.13}$ & $20.82^{+0.31}_{-0.15}$ & $21.45^{+0.06}_{-0.13}$  \\ 
$a_*$ & - & 0.998(f) & (l) & (l) & (l) & (l) & (l) \\
$i$ & deg & $34\pm2$ & (l) &  (l) & (l)  &(l)  & (l) \\
$Z_{\rm Fe}$ & $Z_{\odot}$ & $1.50^{+0.12}_{-0.04}$ & (l) & (l) & (l) & (l) & (l) \\
$\log(F_{\rm refl})$ & erg\,cm$^{-2}$\,s$^{-1}$ & $-9.42^{+0.04}_{-0.09}$ & $-9.72^{+0.07}_{-0.03}$ & $-9.91^{+0.13}_{-0.05}$ & $-10.13^{+0.11}_{-0.09}$ & $-10.21^{+0.03}_{-0.06}$ & $-10.53^{+0.10}_{-0.11}$ \\
\hline
$E_{\rm cut}$ & keV & $350^{+92}_{-124}$  & >287 & >255 & >290 & >350 & >420 \\
$\Gamma$ & - & $1.594^{+0.004}_{-0.010}$ & $1.530^{+0.018}_{-0.045}$ & $1.49^{+0.02}_{-0.05}$ & $1.517^{+0.010}_{-0.030}$ & $1.485^{+0.007}_{-0.029}$ & $1.49^{+0.03}_{-0.02}$ \\
$\log(F_{\rm pl})$ & erg\,cm$^{-2}$\,s$^{-1}$ & $-9.22^{+0.05}_{-0.03}$ & $-9.25\pm0.03$ & $-9.32^{+0.02}_{-0.03}$ & $-9.420^{+0.011}_{-0.028}$ & $-9.601^{+0.009}_{-0.013}$ & $-9.82\pm0.02$ \\
\hline
$kT$ & keV & - & - & - & - & - & - \\
$N_{\rm diskbb}$ & - & - &  - & - & - & - & - \\
\hline
$f_{\rm refl}$ & - & $0.631^{+0.010}_{-0.016}$ & $0.3388^{+0.0041}_{-0.0010}$ & $0.224^{+0.006}_{-0.002}$ & $0.195^{+0.003}_{-0.002}$ & $0.246^{+0.003}_{-0.002}$ & $0.194^{+0.003}_{-0.002}$ \\
$\log(F_{\rm dis})$ & erg\,cm$^{-2}$\,s$^{-1}$ & $-11.01^{+0.14}_{-0.20}$ & $-11.58^{+0.16}_{-0.31}$ & $-11.40^{+0.19}_{-0.33}$ & $-11.4^{+0.2}_{-0.3}$ & $-11.16^{+0.15}_{-0.20}$ & <-11.69 \\
C-stat/$\nu$ & & 11870.25/11305 & & & & & \\
\hline
$L_{\rm 0.1-100keV}/L_{\rm Edd}$ & \% & 2.7 & 2.5 & 2.2 & 1.7 & 1.2 & 0.6 \\
\hline\hline
Continued \\
\hline\hline
Parameter & Unit & LF7 & LF8 & LF9 & LF10 & LF11  & HF \\ 
\hline
$N_{\rm H}$  & 10$^{21}$cm$^{-2}$ & $4.1^{+0.3}_{-0.2}$ & $3.84^{+0.18}_{-0.16}$ & $3.41^{+0.18}_{-0.17}$ & $3.71^{+0.16}_{-0.23}$ & $3.75^{+0.12}_{-0.14}$ & $6.2\pm0.2$ \\
\hline
q & -  & $>0.5$ & $>2$ & >3 &>2.7 & 4(u) & $5.88^{+1.01}_{-0.77}$ \\
rin & ISCO  & <25 & <17 & <10 & <1.51 & <25 & 1(f) \\
$\log(\xi)$ & erg cm s$^{-1}$  & $3.29^{+0.04}_{-0.03}$ & $3.26\pm0.03$ & $3.23^{+0.016}_{-0.020}$ & $3.23^{+0.02}_{-0.05}$ & $3.26^{+0.04}_{-0.03}$ & $3.88^{+0.08}_{-0.12}$ \\
$\log(n_{\rm e})$ & cm$^{-3}$  & $21.0\pm0.2$ &  $21.25^{+0.23}_{-0.17}$ & $21.15\pm0.15$ & $20.93^{+0.12}_{-0.08}$ & $21.57^{+0.20}_{-0.17}$  & $18.93^{+0.12}_{-0.16}$ \\ 
$a_*$ & -  & (l) & (l) & (l) & (l) & (l) & >0.93\\
$i$ & deg &  (l) &  (l) & (l)  &(l)  & (l) & $35.9^{+1.6}_{-2.0}$\\
$Z_{\rm Fe}$ & $Z_{\odot}$ & (l) & (l) & (l) & (l) & (l) & $1.05^{+0.17}_{-0.15}$\\
$\log(F_{\rm refl})$ & erg\,cm$^{-2}$\,s$^{-1}$  & $-10.071^{+0.009}_{-0.010}$ & $-9.93\pm0.04$ & $-9.71^{+0.03}_{-0.02}$ & $-9.52^{+0.03}_{-0.04}$ & $-10.41^{+0.02}_{-0.07}$ & $-8.30^{+0.03}_{-0.02}$ \\
\hline
$E_{\rm cut}$ & keV & >380 & >320 & >430 & >330 & >410 & 500(f) \\
$\Gamma$ & -  & $1.427^{+0.065}_{-0.016}$ & $1.419^{+0.012}_{-0.010}$ & $1.421^{+0.009}_{-0.015}$ & $1.42\pm0.02$ & $1.478^{+0.011}_{-0.018}$ & $2.357^{+0.019}_{-0.018}$ \\
$\log(F_{\rm pl})$ & erg\,cm$^{-2}$\,s$^{-1}$  & $-9.467^{+0.007}_{-0.008}$ & $-9.278^{+0.011}_{-0.013}$ & $-9.068^{+0.010}_{-0.021}$ & $-8.94^{+0.009}_{-0.008}$ & $-9.701^{+0.015}_{-0.014}$ & $-8.46^{+0.06}_{-0.05}$ \\
\hline
$kT$ & keV &  - & - & - & - & - & $0.831^{+0.03}_{-0.05}$ \\
$N_{\rm diskbb}$ &  - &  - & - & - & - & - & $1649^{+76}_{-49}$\\
\hline
$f_{\rm refl}$ & -  & $0.249^{+0.005}_{-0.007}$ & $0.22^{+0.02}_{-0.02}$ & $0.228^{+0.017}_{-0.011}$ & $0.26\pm0.02$ & $0.195^{+0.012}_{-0.030}$ & $1.45^{+0.06}_{-0.03}$ \\
$\log(F_{\rm dis})$ & erg\,cm$^{-2}$\,s$^{-1}$ &  $-11.00^{+0.10}_{-0.12}$ & $-10.73^{+0.02}_{-0.13}$ & $-10.62\pm0.08$ & $-10.50^{+0.09}_{-0.08}$ & $-1.06^{+0.08}_{-0.13}$ & <-10.89\\
C-stat/$\nu$ & & & & & & & 1048.68/971 \\
\hline
$L_{\rm 0.1-100keV}/L_{\rm Edd}$ & \% & 1.2 & 2.0 & 3.1 & 4.0 & 0.7 & 26.7 \\
\hline\hline
\end{tabular}
\label{tab_fit}
\end{table*}


\subsection{High Flux State (HF) Spectral Modelling} \label{hf_spectrum}

The same \nustar and \swift observations of GX~339-4 in a HF state analysed in \citet{parker16} are considered here. A ratio plot of the HF spectra fitted with a Galactic absorbed power-law model and a simple disc blackbody component \texttt{diskbb} is shown in the bottom panel of Fig.\,\ref{pic_fe}. The HF spectra show a broad emission line feature at the iron band and a Compton hump above 10~keV, indicating existence of a relativistic disc reflection component similar with all the LF spectra. A multicolour disc blackbody component \texttt{diskbb} is used to account for the strong disc thermal component. The full model is \texttt{tbabs * ( cflux*(relconv*reflionx) + cflux*reflionx + cflux*cutoffpl + diskbb)} in XSPEC format. This model provides a good fit with C-stat/$\nu$=1048.68/971. The best-fit model is shown in the last panel of Fig.\,\ref{pic_final} and the best-fit parameters are shown in the last column of Table\,\ref{tab_fit}. A disc density of $\log(n_{\rm e}/$cm$^{-3})=18.93^{+0.12}_{-0.16}$ is found in HF observations which is 100 times lower than the best-fit value in LF observations.

So far we have obtained a good fit for the HF spectrum of GX~339-4. A higher column density is required for the neutral absorber ($N_{\rm H}=6.2\pm0.2\times10^{21}$\,cm$^{-2}$) compared to the LF observations ($N_{\rm H}\approx3.4\times10^{21}$\,cm$^{-2}$). \citet{parker16} obtained a higher column density of $N_{\rm H}=7.2\pm0.2\times10^{21}$\,cm$^{-2}$ for the same observation assuming $n_{\rm e}=10^{15}$cm$^{-3}$ for the disc. Both our result and \citet{parker16} are higher than the Galactic absorption column density estimated at other wavelengths \citep[e.g.][]{kalberla05}, indicating a possible extra variable neutral absorber along the line of sight. Only an upper limit of the flux of the distant cold reflector is achieved ($\log(F_{\rm dis})<-10.89$). The 1--10~keV band flux values of the disc reflection component and the coronal emission increase by a factor of 13 and 6 respectively compared to LF1. The best-fit broad band model shows the highest reflection fraction among all the observations considered in this work, indicating a geometry change of the disc corona system such as a small inner disc radius. A solar iron abundance ($1.05^{+0.17}_{-0.15}$) is required for the HF spectra, which is lower than the value obtained by \citet{parker16}, where a disc density of $n_{\rm e}=10^{15}$cm$^{-3}$ is assumed.

\section{Results and Discussion} \label{results_spectrum}

We have obtained a good fit for the LF and the HF pectra of GX~339-4. The LF spectral modelling requires a high disc density of $\log(n_{\rm e}/{\rm cm^{-3}})\approx21$ with no additional low temperature thermal component. The HF spectral modelling requires a 100 times lower density ($\log(n_{\rm e}/{\rm cm^{-3}})=18.93^{+0.12}_{-0.16}$) compared to LF observations. In this section, we discuss the spectral fitting results by comparing with previous data analysis and accretion disc theories.

\subsection{Comparison with previous results}

\begin{figure*}
    \centering
	\includegraphics[width=15cm]{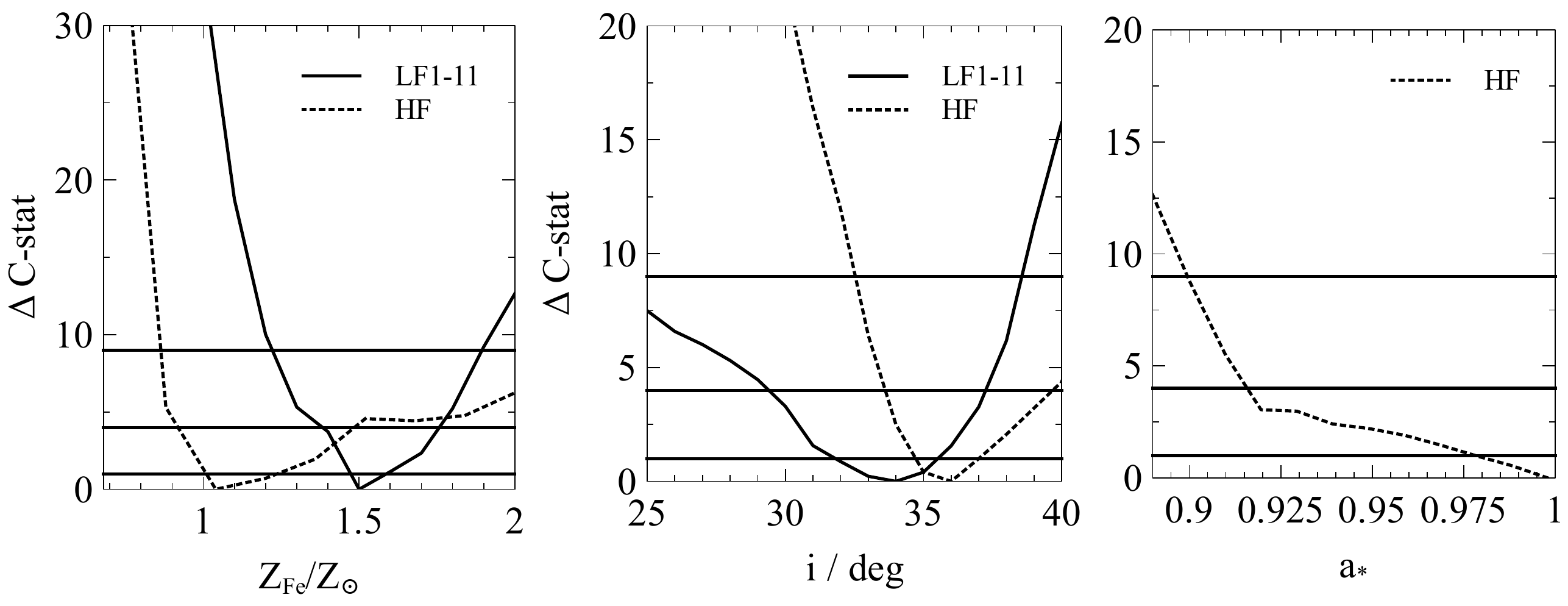}
    \caption{C-stat contour plots for the disc iron abundance and the relativistic reflection parameters obtained by fitting LF 1-11 spectra simultaneously (solid lines) and only the HF spectra (dashed line). The solid lines show the 1$\sigma$, 2$\sigma$, and 3$\sigma$ ranges.}
    \label{pic_contour_all}
\end{figure*}

First, the most significant difference from previous results is the close-to-solar disc iron abundance. Previously, \citet{parker16,jingyi18} obtained a disc iron abundance of $Z_{\rm Fe}=6.6^{+0.5}_{-0.6}Z_{\odot}$ and $Z_{\rm Fe} \approx 8 Z_{\odot}$ respectively by analysing the same spectra considered in this work. Similar result was achieved by \citet{furst15}. A high iron abundance of $Z_{\rm Fe}=5\pm1Z_{\odot}$ was also found by analysing stacked \textit{RXTE} spectra \citep{garcia15}. All of their work was based on the assumption for a fixed disc density $n_{\rm e}=10^{15}$\,cm$^{-3}$. By allowing the disc density $n_{\rm e}$ to vary as a free parameter during spectral fitting, we obtained a close-to-solar disc iron abundance ($Z_{\rm Fe}=1.50^{+0.12}_{-0.04}Z_{\odot}$ for LF observations and $Z_{\rm Fe}=1.05^{+0.17}_{-0.15}Z_{\odot}$ for HF observations). The best-fit disc iron abundance for the LF spectra is slightly higher than the value for the HF spectra at 90\% confidence. However they are consistent within 2$\sigma$ confidence range. See the left panel of Fig.\,\ref{pic_contour_all} for the constraints on the disc iron abundance parameter. A similar conclusion was achieved by analysing the intermediate flux state spectra of Cyg X-1 \citep{tomsick18} with variable density reflection model. However a fixed solar iron abundance was assumed in their modelling.

Second, the best-fit disc viewing angle measured for GX~339-4 is different in different works. The middle panel of Fig.\,\ref{pic_contour_all} shows the constraint of the disc viewing angle given by multi-epoch LF spectral analysis and HF spectral analysis. \blue{The} two measurements are consistent at the 90\% confidence level ($i=34\pm2$\,deg for the LF observations and $i=35.9^{+1.6}_{-2.0}$\,deg for the HF observations). Although our best-fit value is lower compared with the measurement in \citet{jingyi18} ($i=40$$^{\circ}$) and higher than the measurement in \citet{parker16} ($i=30$$^{\circ}$), all the previous reflection based measurements are consistent with our results at 2$\sigma$ level. Similarly \citet{tomsick18} measured a different viewing angle for Cyg X-1 compared with previous works. It indicates that a slightly different viewing angle measurement might be common when allowing the disc density $n_{\rm e}$ to vary as a free parameter during the spectral fitting.

Third, a high black hole spin ($a_*>0.93$) is given by the disc reflection modelling in the HF spectral analysis. Due to the lack of precise measurement of the source distance and the central black hole mass, we can only give a rough estimation of the inner radius through the normalization of the \texttt{diskbb} component in the HF observations. The normalization parameter is defined as $N_{\rm diskbb}=(r_{\rm in}/D_{\rm10kpc})^{2}\cos i$, where the $D_{\rm10kpc}$ is the source distance in units of 10\,kpc and $i$ is the disc viewing angle. The best-fit value is $N_{\rm diskbb}=1649^{+76}_{-49}$, corresponding to an inner radius of $r_{\rm in}\approx45$\,km$\approx3r_g$ assuming $M_{\rm BH}=10M_{\odot}$ and $D=10$\,kpc. We also fitted the thermal component with \texttt{kerrbb} model \citep{li05} as in \citet{parker16}. \texttt{kerrbb} is a multi-colour blackbody model for a thin disc around a Kerr black hole, which includes the relativistic effects of spinning black hole. The BH spin and the disc viewing angle are linked to the corresponding parameters in \texttt{relconv}. However we found the source distance and the central black hole mass parameters in \texttt{kerrbb} are not constrained during the spectral fitting. \texttt{kerrbb} model gives a slightly worse fit with $\Delta$C-stat$\approx7$ and 2 more free parameters compared to the \texttt{diskbb} model. Since the black hole mass and distance measurement is beyond our purpose, we decide to fit the thermal spectrum in the HF observation with \texttt{diskbb} for simplicity. See \citet{parker16} for more discussion concerning the black hole mass and the source distance measurements obtained by fitting with \texttt{kerrbb}. In conclusion, the high spin result in this work is obtained by modelling the relativistic disc reflection component in the HF state of GX~339-4 and consistent with previous reflection-based spectral analysis \citep[e.g.][]{plant15, garcia15, parker16, jingyi18}. \citet{kolehmainen10} found an upper limit of $a_*<0.9$ by analysing \textit{RXTE} spectra. However they assumed the disc viewing angle is the same as the binary orbital inclination, which is not necessarily the case \citep[e.g.][]{tomsick14,walton16}.

\begin{figure*}
    \centering
	\includegraphics[width=13cm]{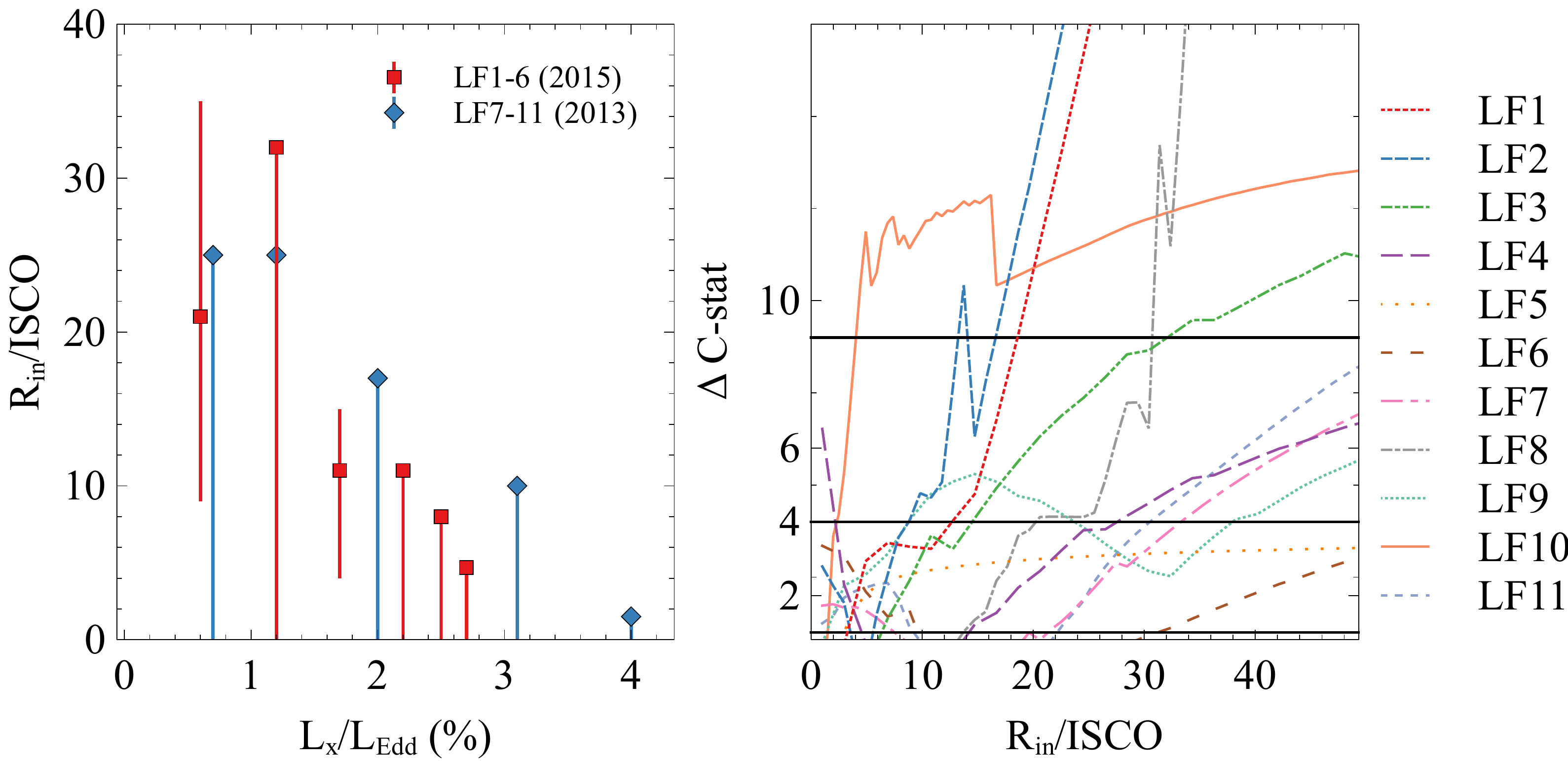}
    \caption{Left: the best-fit inner radius of the disc vs. the X-ray luminosities for LF observations. Red points represent LF1-6 observations and blue points represent LF7-11 observations. The X-ray luminosity $L_{\rm X}$ is the 0.1--100~keV band Galactic-absorption corrected luminosity calculated using the best-fit model. A black hole mass $M_{\rm BH}=10M_{\odot}$ and a distance $d$=10\,kpc are assumed. The error bars show the 90\% confidence ranges of the measurements. See Table\,\ref{tab_fit} for values. Right: the constraints on the inner disc radius for each LF spectra shown in different colours. The solid lines show the 1$\sigma$, 2$\sigma$, and 3$\sigma$ ranges.}
    \label{pic_contour_rin}
\end{figure*}

Fourth, there is a debate whether the disc is truncated at a significant radius in the brighter phases of the hard state. Compared with the results obtained by modelling the same spectra with $n_{\rm e}=10^{15}$cm$^{-3}$ and an additional \texttt{diskbb} component \citep{jingyi18}, we find larger upper limit of the inner radius in the LF2-5 observations. For example, an upper limit of $r_{\rm in}<8R_{\rm ISCO}$ is obtained for the LF2 observation, larger than $r_{\rm in}=1.8^{+3.0}_{-0.6}R_{\rm ISCO}$ found by \citet{jingyi18}. Such difference could be due to different modelling of the disc reflection component. The constraints on the inner radius $r_{\rm in}$ are shown in the top right panel of Fig.\,\ref{pic_contour_rin}.  Nevertheless, we confirm that the inner radius is not as large as $r_{\rm in}\approx 100r_g$ as proposed by previous analysis \citep[e.g.][]{plant15}.

\subsection{High density disc reflection}

The LF and HF \nustar and \swift spectra of GX~339-4 can be successfully explained by high density disc reflection model with a close-to-solar iron abundance for the disc. In the low flux hard states, no additional low-temperature \texttt{diskbb} component is required in our modelling. Instead, a  quasi-blackbody emission in the soft band of the disc reflection model fits the excess below 2~keV. At higher disc density, the free-free process becomes more important in constraining low energy photons, increasing the disc surface temperature, and thus turning the reflected emission in the soft band into a quasi-blackbody spectrum. See Fig.\,\ref{pic_comparison} for a comparison between the best-fit high density reflection model for LF1 observation and a constant disc density model ($n_{\rm e}=10^{15}$\,cm$^{-3}$).

In LF states of GX~339-4, a disc density of $n_{\rm e}\approx10^{21}$\,cm$^{-3}$ is required for the spectral fitting. Our multi-epoch spectral analysis shows tentative evidence that the disc density increases from $\log(n_{\rm e}/{\rm cm^{-3}})=20.60^{+0.23}_{-0.12}$ in the highest flux state (LF1) to $\log(n_{\rm e}/{\rm cm^{-3}})=21.45^{+0.06}_{-0.13}$ in the lowest flux state (LF6) during the decay of the outburst in 2015, except for LF5 observation. See Table\,\ref{tab_fit} for $n_{\rm e}$ measurements. Similar pattern can be found in LF7-10 observations. In HF state of GX~339-4, we measure a disc density of $n_{\rm e}\approx10^{19}$\,cm$^{-3}$ by fitting the broad band spectra with a combination of high density disc model and a multi-colour disc blackbody model. The disc density in HF state is 100 times lower than that in LF states. The 0.1--100~keV band luminosity of GX~339-4 in HF state ($L_{\rm X}\approx0.28L_{\rm Edd}$) is 10 times the same band luminosity in LF states ($L_{\rm X}=0.01-0.03L_{\rm Edd}$). While the accretion rate is rather small, the anti-correlation between the disc density and the X-ray luminosity $\log(n_{\rm e1}/n_{\rm e2})\propto-2\log(L_{\rm X1}/L_{\rm X2})$ is found to agree with the expected behaviour of a standard radiation-pressure dominated disc \citep[e.g. ][]{shakura73,svensson94}. See Section \ref{appen1} for more detailed comparison between the measurements of the disc density and the predictions of the standard disc model.

\subsection{Accretion rate and disc density} \label{appen1}

\begin{figure}
	\includegraphics[width=\columnwidth]{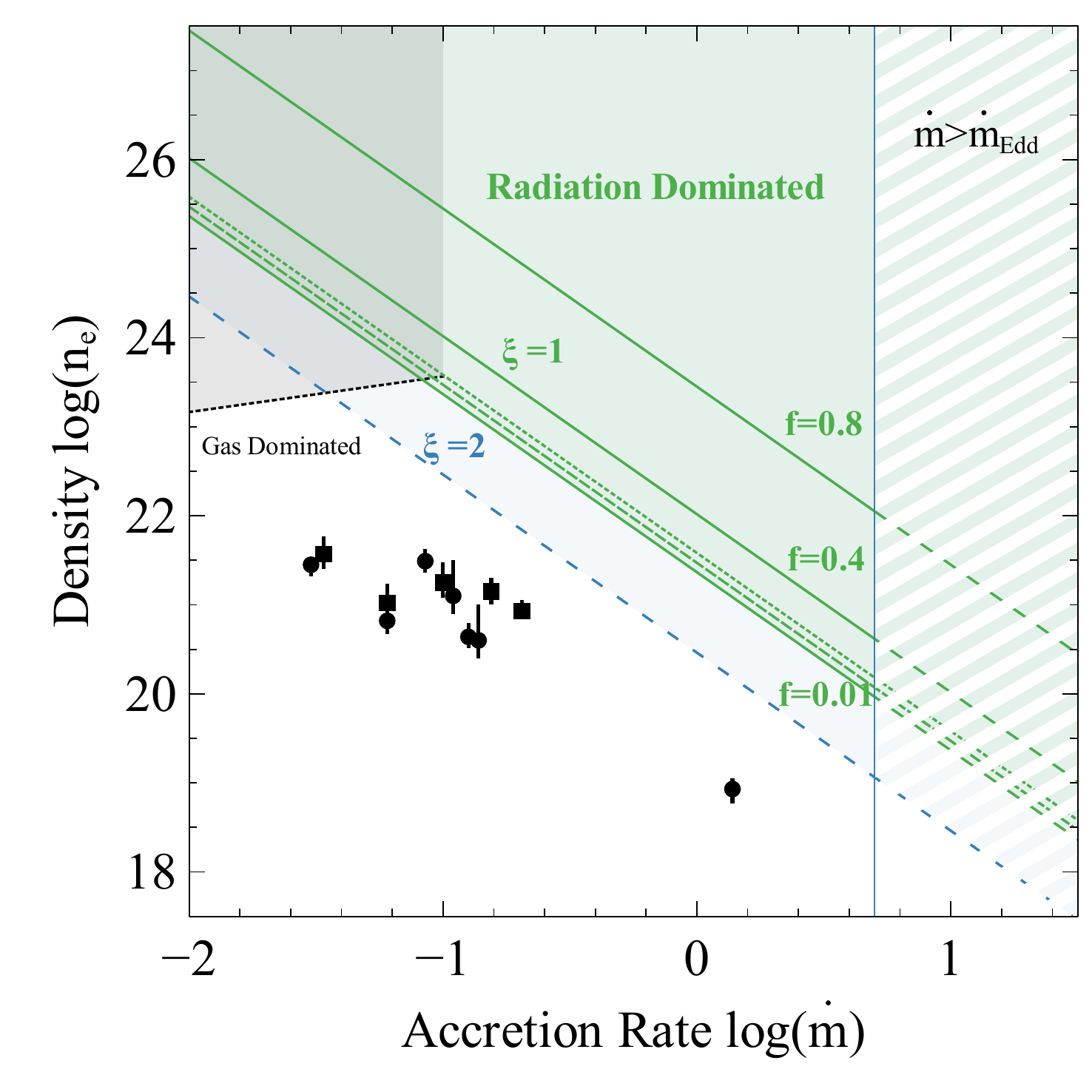}
    \caption{Disc density $\log(n_{\rm e}/{\rm cm^{-3}})$ vs. accretion rate $\log(\dot{m})$ based on the radiation-pressure dominated (green) and gas-pressure dominated (black) disc solutions in SZ94, assuming $M_{\rm BH}=10M_{\odot}$ and $\alpha=0.1$. The solid green lines show the solutions with different coronal power fraction $f$ at $R=2R_{\rm S}$. The dashed and dotted green lines show the radiation pressure-dominated solution with $f=0.01$ at $R=6,8$$R_{\rm S}$ respectively. The black circular points show the surface disc density and the mass accretion rate measurements of GX~339-4 in LF and HF states in 2015 and the black squares show the measurements for observations in 2013. The mass accretion rate is estimated using $\dot{m}=L_{\rm Bol}/\epsilon L_{\rm Edd}\approx L_{\rm 0.1-100keV}/\epsilon L_{\rm Edd}$ in this work. A Novikov-Thorne accretion efficiency $\epsilon=0.2$ \citep{novikov73,agol00} and an inner disc radius of $R_{\rm in}=1R_{\rm S}$ is assumed for a spinning black hole with $a_*=0.95$. The black vertical line shows the Eddington accretion limit $\dot{m}_{\rm Edd}=1/\epsilon\approx5$. $\xi=1$ is assumed during the calculation as in SZ94. The blue solid line is shown for the radiation-pressure dominated disc solution with $R=2R_{\rm S}$ and $f=0.01$, assuming $\xi=2$.} 
    \label{pic_ne_f}
\end{figure}

\citet{svensson94} (hereafter SZ94) reconsidered the standard accretion disc model \citep[][hereafter SS73]{shakura73} by adding one more parameter to the disc energy balance condition - a fraction of the power associated with the angular momentum transport is released from the disc to the corona, denoted as $f$. Only $1-f$ of the released accretion power is dissipated in the colder disc itself. 
 
By following SZ94, we can obtain a relation between $n_{\rm e}$ and $f$ for a radiation-pressure dominated disc:
\begin{equation}
n_{\rm e} = \frac{1}{\sigma_{\rm T}R_{\rm S}}\frac{256\sqrt[]{2}}{27} \alpha^{-1}R^{3/2}\dot{m}^{-2}[1-(R_{\rm in}/R)^{1/2}]^{-2}[\xi(1-f)]^{-3},
\label{eq1}
\end{equation}
where $\sigma_{\rm T}=6.64\times10^{25}$cm$^{2}$ is the Thomson cross section; $R_{S}$ is the Schwarzschild radius; $R$ is in the units of $R_{\rm S}$; $\dot{m}$ is defined as $\dot{m}=\dot{M}c^{2}/L_{\rm Edd}=L_{\rm Bol}/\epsilon L_{\rm Edd}$; $L_{\rm Edd}=4\pi GMm_{\rm p}c/\sigma_{\rm T}=2\pi (m_{\rm p}/m_{\rm e})(m_{\rm e}c^3/\sigma_{\rm T})R_{\rm S}$ is the Eddington luminosity; $\xi$ is the conversion factor in the radiative diffusion equation and chosen to be 1, 2 or 2/3 by different authors (SZ94). For a radiation-pressure dominated disc, the disc density $n_{\rm e}$ decreases with increasing accretion rate $\dot m$. In Fig.\,\ref{pic_ne_f}, we plot the radiation-pressure dominated disc solutions for $\xi=1$ in green lines and a solution for $\xi=2$ in blue.  

When $\dot m<0.1$ and $f$ approaches unity, the radiation-pressure dominated radius disappears and gas pressure starts dominates the disc. The relation between $n_{\rm e}$ and $f$ for a gas-pressure dominated disc is
\begin{equation}
n_{\rm e} = \frac{1}{\sigma_{\rm T} R_{\rm S}} K \alpha^{-7/10} R^{-33/20} \dot{m}^{2/5} [1-(R_{\rm in}/R)^{1/2}]^{2/5} [\xi(1-f)]^{-3/10},
\label{eq2}
\end{equation}
where $K=2^{-7/2} (\frac{512\sqrt{2}\pi^3}{405})^{3/10}(\alpha_{\rm f}\frac{m_{\rm p}}{m_{\rm e}})^{9/10} (\frac{R_{\rm S}}{r_{\rm e}})^{3/10}$. $\alpha_{\rm f}$ is the fine-structure constant; $m_{\rm p}$ is the proton mass; $m_{\rm e}$ is the electron mass; $r_{\rm e}$ is the classical electron radius. An example of a gas-pressure dominated disc solution for $R=2R_{\rm S}$ and $f=0.01$ is shown by the black line in Fig.\,\ref{pic_ne_f} for comparison. For a gas-pressure dominated disc, the disc density increases with increasing accretion rate.

The best-fit disc density and accretion rate values obtained by fitting GX~339-4 LF1-11 and HF spectra with high density reflection model are shown by black points in Fig.\,\ref{pic_ne_f}. The accretion rate is calculated using $\dot{m}=L_{\rm Bol}/\epsilon L_{\rm Edd}\approx L_{\rm 0.1-100keV}/\epsilon L_{\rm Edd}$, where $\epsilon$ is the accretion efficiency and $L_{\rm 0.1-100keV}$ is the 0.1--100keV band absorption corrected luminosity calculated using the best-fit model.  According to \citet{novikov73,agol00}, an accretion efficiency of $\epsilon=20\%$ is assumed for a spinning black hole with $a_*=0.95$ measured in Section\,\ref{hf_spectrum}. A black hole mass  of 10$M_{\odot}$ and a source distance of 10\,kpc are assumed. 

The 0.1--100~keV band luminosity in the HF state of GX~339-4 is approximately 10 times the same band luminosity in the LF states. The disc density in the HF state is 2 orders of magnitude lower than the density in the LF1-6 states. The anti-correlation between its density and accretion rate is expected according to the radiation-pressure dominated disc solution in SZ94 ($\log(n_{\rm e})\propto -2 \log(\dot m)$, as in Eq.\,\ref{eq1}). However the disc density measurements for GX~339-4 are lower than the predicted values for corresponding accretion rates. See Fig.\,\ref{pic_ne_f} for comparison between measurements and theoretical predictions in SZ94. Following are possible explanations for the mismatch: 1. the disc density shown in Eq.\,\ref{eq1} is assumed to be constant throughout the vertical profile of the disc (SS73). The $n_{\rm e}$ parameter we measure using reflection spectroscopy is however the surface disc density within a small optical depth \citep{ross07}. For example, three-dimensional MHD simulations show that the vertical structure of radiation-pressure dominated disc density is centrally concentrated \citep[e.g.][]{turner04,hirose06}. 2. the accretion rate might be underestimated by assuming $L_{\rm Bol}\approx L_{\rm 0.1-100keV}$, although we do not expect other bands of GX~339-4 to make a large contribution to its bolometric luminosity; 3. there is a large uncertainty on the black hole mass, the disc accretion efficiency, and the source distance measurements for GX~339-4. For example, the most recent near-infrared study shows that the central black hole mass in GX~339-4 could be within  $2-10M_{\odot}$ \citep{heida17}. Nevertheless, the use of the high density reflection model enables us to estimate the density of the disc surface in different flux states of an XRB and an anti-correlation between $n_{\rm e}$ and $L_{\rm X}$ has been found in GX~339-4. 

\subsection{Future work}

In our work, we conclude that the high density reflection model can explain both the LF and HF spectra of GX~339-4 with a close to solar iron abundance. No additional blackbody component is statically required during the spectral fitting of the LF states. On one hand, the strong degeneracy between the \texttt{diskbb} component and the absorber column density is due to the low S/N of the \swift XRT observations. More pile-up free soft band spectra are required to obtain a more detailed spectral shape at the extremely LF state of GX~339-4, such as from \textit{NICER}. On the other hand, more detailed spectral modelling is required. For example, a more physics model, such as Comptonization model, is required for the coronal emission modelling in the broad band spectral analysis. The disc thermal photons from the disc to the reflection layer need to be taken into account in future reflection modelling, especially in the XRB soft states where a strong thermal spectrum is shown.

\section*{Acknowledgements}

J.J. acknowledges support by the Cambridge Trust and the Chinese Scholarship Council Joint Scholarship Programme (201604100032). D.J.W. acknowledges support from an STFC Ernest Rutherford Fellowship. A.C.F. acknowledges support by the ERC Advanced Grant 340442. M.L.P. is supported by European Space Agency (ESA) Research
Fellowships. J.F.S. has been supported by NASA Einstein Fellowship grant No. PF5-160144. J.A.G. acknowledges support from the Alexander von
Humboldt Foundation. This work made use of data from the \nustar mission, a project led by the California Institute of Technology, managed by the Jet Propulsion Laboratory, and funded by NASA. This research has made use of the \nustar Data Analysis Software (NuSTARDAS) jointly developed by the ASI Science Data Center and the California Institute of Technology. This work made use of data supplied by the UK Swift Science Data Centre at the University of Leicester. We acknowledge support from European Space Astronomy Center (ESAC).




\bibliographystyle{mnras}
\bibliography{gx33942.bib} 




\appendix


\bsp	
\label{lastpage}
\end{document}